\documentclass[preprint, authoryear, 5p, twocolumn]{elsarticle}

\usepackage[utf8]{inputenc} 
\usepackage[T1,T2A]{fontenc}    
\usepackage[english]{babel} 
\usepackage{hyperref} 
\usepackage{indentfirst}
\usepackage{graphicx}
\usepackage{epstopdf}
\usepackage{url}            
\usepackage{booktabs}       
\usepackage{amsfonts}       
\usepackage{nicefrac}       
\usepackage{microtype}      
\usepackage{lipsum}
\usepackage{xcolor}
\usepackage{subcaption}
\usepackage{environ}
\usepackage{pifont} 
\usepackage{enumitem}
\usepackage{comment}
\usepackage{amsmath, amsfonts, amssymb}
\usepackage{tabularx,ragged2e,booktabs,caption} 
\newcolumntype{C}[1]{>{\Centering}m{#1}}

\usepackage{titlesec}
\usepackage{mdframed}
\usepackage[font=small,labelfont=bf,textfont=normalfont]{caption}
\usepackage{placeins}    
\usepackage{xurl}

\newcommand{\band}[1]{\mathcal{I}_{#1}}
\newcommand{\nubeg}[1]{\nu_{#1}^{-}}
\newcommand{\nuend}[1]{\nu_{#1}^{+}}

\usepackage{listings}

\let\parencite\citep
\let\textcite\citet

\begin{document}
\begin{frontmatter}
    \title{Vertically resolved minimal-set k-distribution for thermal infrared absorption: an application to the atmosphere of Venus}
    \author[1]{Boris Fomin}
    \author[2]{Mikhail Razumovskiy\corref{cor1}}
    \ead{razumovskii@phystech.edu}
\cortext[cor1]{Corresponding author}
\affiliation[1]{organization={Central Aerological Observatory},
            addressline={3 Pervomayskaya St.}, 
            city={Dolgoprudny},
            postcode={141700}, 
            state={Moscow Region},
            country={Russia}}

\affiliation[2]{organization={Moscow Institute of Physics and Technology},
            addressline={9 Institutskiy per.}, 
            city={Dolgoprudny},
            postcode={141701}, 
            state={Moscow Region},
            country={Russia}}

\begin{abstract}
The FKDM $k$-distribution technique is applied to parameterize absorption of thermal radiation in the lower and middle atmosphere of Venus, targeting modeling scenarios, where the cost of full radiative transfer calculations necessitates efficient parameterizations (e.g. climate modeling). Line-by-line reference modeling based on a Monte Carlo method for radiative transfer is built into the $k$-distribution terms construction process, explicitly controlling accuracy. From 16 bands across 10–6000 cm$^{-1}$, the method produces 32 $k$-terms, band-averaged Planck function values and per-band spectral points for computing Venus cloud optical properties. The FKDM $k$-distribution technique does not require the inter-level correlation assumption common for the correlated $k$-distribution method. We supply height-dependent $k(z)$ functions tabulated on the same vertical grid as the input temperature-pressure profile, designed for direct use in radiative transfer solvers and avoiding additional remapping of the pre-tabulated $k$-data. Our implementation of the technique yielded acceptable accuracy below 90 km (<1.2 $\mathrm{K\,day^{-1}}$ for cooling rates; <2~\% for fluxes), while requiring substantially fewer $k$-terms than recent implementations of the correlated-$k$ method. A Fortran driver that generates $k(z)$ functions for arbitrary Venus atmospheric profile is provided in a public repository.
\end{abstract}

    

\end{frontmatter}
\section{Introduction}
\label{sec:intro}

Thorough consideration of all aspects of radiative transfer in Venus climate modeling still represents a substantial challenge. The problem domain can be considered two-fold. Firstly, despite recent attempts for exoplanets \parencite{ding2019new}, for Venus — heavy reference line-by-line calculations are presently considered unrealistic to integrate especially into time-marching models such as general circulation models (GCM) or radiative-convective equilibrium models (RCM). Therefore, radiation schemes must be introduced in a parameterized form, and development of such parameterizations stands as an independent research issue. Secondly, accurate reference radiative transfer modeling is still required to validate parameterizations against line-by-line results. However, current uncertainties in Venus atmospheric and spectroscopic parameters (e.g. deep atmospheric opacities, profiles of minor species) limit the accuracy and reproducibility of the reference modeling itself.

Accurate treatment of molecular absorption is central to radiative transfer schemes, yet its parameterization is indispensable due to the vast number of individual spectral lines to be involved. Line-by-line calculation of gaseous absorption properties from molecular line lists also demands substantial computational resources. Since pioneering studies by \textcite{goody1989correlated, LacisOinas1991, fu1992correlated}, the correlated $k$-distribution has become a principal technique for parameterizing gaseous absorption and reducing the number of radiative transfer equation evaluations. This method has been extensively employed in Venus radiative-convective modeling \parencite{bullock1996stability, bullock2001recent}, steady-state radiative transfer models (e.g., \textcite{Tsang2008, lee2011discrete}), and energy balance studies \parencite{titov2007radiation, lebonnois2015analysis}. It has also been applied in general-purpose codes for spectral retrievals and for radiative transfer modeling applicable to Venus \parencite{irwin2008nemesis, villanueva2018planetary, takahashi2023development}. It also serves as the foundation for radiative transfer modules for three-dimensional general circulation models with higher levels of parameterization over broad spectral ranges: \textcite{eymet2009net}, \textcite{mendonca2015new}, and \textcite{tahseen2024enhancing}.

Correlated-$k$ approach assumes correlation of spectral ordering between atmospheric layers, which introduces a systematic error, generally considered acceptable for Earth atmosphere modeling \parencite{liou2002introduction}. The accuracy implications of this assumption are usually mitigated by introducing more bands and/or quadrature points within the band. However, the trade-off between overall accuracy and the number of $k$-distribution terms remains an area for improvement under various atmospheric conditions \parencite{cusack1999investigating, hogan2010full, pincus2019balancing}. This is particular for Venus, where vertical atmospheric structure features high pressure-temperature gradients and deep high-temperature conditions that introduce additional spectroscopic uncertainties (e.g., continuum absorption), potentially amplifying the correlation assumption errors. Moreover, the required precision for cooling and heating rates in Venus GCMs—to resolve phenomena like mesospheric dynamics—remains an ongoing challenge and not yet fully established. Overall, computational costs of radiative transfer equation evaluations remain high with correlated-$k$-based parameterizations; for instance, \textcite{mendonca2015new} preferred band-averaged absorption values over the correlated-k approach when developing the radiative transfer scheme for the Oxford Venus GCM. Though today correlated-$k$ method is instrumental for Venus atmosphere modeling, to our knowledge, techniques alternative to correlated-$k$, such as FSCK method \parencite{hogan2010full} or l-distributions method \parencite{tapimo2024fast} have not been applied for describing Venus thermal infrared absorption yet. Recently, along a different line, machine-learned surrogate radiative transfer models have been integrated into the OASIS Venus GCM to accelerate the RT module, reporting substantial speed-ups \parencite{tahseen2024enhancing}.

In this study we implement the FKDM $k$-distribution technique, initially developed by \textcite{Fomin2004} and successfully implemented for Earth atmosphere \parencite{Fomin2004, Fomin2005}. The method utilizes line-by-line reference fluxes and cooling rates as organizing principle, enabling a substantial reduction in k-terms and consequent radiative transfer evaluations. This approach also offers potentially higher accuracy, natively accounts for atmospheric vertical inhomogeneity and multi-gas line overlaps. Recently we have already approbated this approach for Venus conditions to obtain $k$-distribution terms for UV absorption in \textcite{Fomin2022}. In the present study we demonstrate the adaptation of the same technique for thermal infrared radiation absorption.

In our workflow, spectral absorption coefficients are precomputed and used both to generate vertically resolved $k$-distribution terms (see step~2 in the vertical iterative procedure, subsection~\ref{subsec:iterative procedure}) and as input for reference radiative transfer modeling. In the correlated-$k$ method, these high-resolution absorption spectra likewise serve as the starting point for the sorting procedure. Consequently, accurate calculation of the line-by-line cross-sections is critical for achieving reliable results. A variety of open-source tools exist for computing absorption and emission spectra under different $P$--$T$ conditions, such as the LBLRTM framework~\parencite{LBLRTM1992}, \textsc{Exocross}~\parencite{yurchenko2018exocross}, and \textit{kspectrum}~\parencite{eymet2016kspectrum}. In this study, however, we employ our recently developed MARFA code~\parencite{Razumovskiy2025MARFA} as the line-by-line backend. For Venusian conditions, uncertainties in spectroscopic parameters --- such as line parameters, line wing treatments, and continuum prescriptions --- necessitate a tool capable of rapidly and efficiently (re)computing absorption spectra under updated assumptions. MARFA is a line-by-line Fortran computing core built on an efficient eleven-grid interpolation scheme~\parencite{fomin1995effective} (compared with the typical three-grid approach), featuring high-resolution spectral calculations of \(1/2048~\mathrm{cm^{-1}}\). The output structure of MARFA differs from the correlated-$k$ table format employed, for example, by \textcite{chaverot2025spect}, but serves the requirements of our workflow. MARFA is suitable for handling large line cut-offs, which are employed in this work together with the $\chi$-factor formalism. Further details are provided in subsection~\ref{sec:absorption by gaseous}.

In Section 2 we provide more details of the FKDM $k$-distribution technique and comparison of the cooling rates obtained from the fast method with those from reference calculations. Section 3 describes modeling inputs for the reference line-by-line model, with particular emphasis on molecular absorption. Section 4 is dedicated to explanation of data/codes provision and how to recalculate parameterizations for a given temperature profile.

\section{$K$-distribution technique}


We follow the fast $k$-distribution technique (FKDM) by \textcite{Fomin2004} and apply it to the Venus environment. Instead of assuming inter-layer correlation of spectrally ordered absorption coefficients, FKDM retrieves a compact set of effective, height-dependent $k$-distribution terms $k(z)$ within each spectral band through an explicit vertical inversion procedure. The resulting $k(z)$ functions are constructed directly on the model’s vertical grid and constrained by radiative fluxes and cooling rates at each level. At the cost of performing repeated comparisons with line-by-line reference calculations, the proposed technique allows full control over the accuracy-efficiency trade-off, producing the smallest number of $k$-distribution terms consistent with a target precision.

For convenience, the technique’s flow can be represented as two consecutive parts. In first part some preliminary actions are made including division the initial spectral range to bands. In the second part, the vertical iterative procedure is employed to construct height-dependent k-terms $k(z)$ within each band. We outline the main steps of the method here, but for a more detailed general description of the approach, please refer to the \textcite{Fomin2004}.

\subsection{Preparatory phase}
\begin{enumerate}[label=\roman*.]
    \item From the molecular line lists, spectral absorption coefficients must be pre-calculated for all optically active gaseous components in the atmosphere. These are computed at each atmospheric level, accounting for temperature dependence, and prepared in a format suitable for direct use in a radiative transfer scheme. We assume temperature and pressure are set constant per layer. For more details on our implementation, refer to subsection \ref{sec:absorption by gaseous}

    \item Initial spectral range (in this study, 10—6000 cm$^{-1}$) must be divided into $n$ bands $\band{i}=[\nubeg{i},\nuend{i}]$, $i = 1,\ldots,n$. Since we concentrate on absorption only by gaseous species, the main criterion for forming bands was the assumption that cloud optical properties are nearly constant within the band. To estimate that, multiple reference calculations were performed for constant and wavelength-dependent cloud optical properties. We have independently constructed the bands this way and found that they are mostly similar to those from the recent study by \textcite{takahashi2023development}.

    \item The target precision must be defined to determine when the k-term generation procedure is terminated, based on the deviation of fast $k$-distribution calculations from the corresponding line-by-line reference results. To our knowledge, no widely accepted precision benchmarks have been established in Venus climate modeling. In the study by \textcite{mendonca2015new}, the band-averaged calculation produces errors in cooling rates in the cloud region not exceeding 10\% deviation, which was considered sufficient for their fast calculation. We follow the accuracy figure of $2 \times 10^{-4}$ $\mathrm{W\,m^{-3}}$ for flux convergence suggested by \textcite{takahashi2023development}, which roughly corresponds to 1.8 $\mathrm{K\,day^{-1}}$ at 80 km. 

    \item To retrieve the absorption profile by vertical iteration, one hemispheric flux must be designated as the reference. We adopt the downward flux and perform a top-down inversion with the boundary condition $F_{\downarrow}(z_{\text{top}})=0$, following the approach of \textcite{Fomin2004}. Although the scheme is formally symmetric with respect to the upward flux $F_{U}$, a bottom-up inversion is ill-conditioned for Venus because, in the optically thick sub-cloud longwave regime, the radiation field is close to Planckian equilibrium and $F_{\downarrow} \approx F_{\uparrow}$, which provides poor numerical leverage.

    \item The vertical interpolation technique is applied individually to each species. Following Venus infrared energetics assessments by \textcite{Haus2015}, only the absorption of CO$_2$, H$_2$O, and SO$_2$ is considered.

\end{enumerate}

\subsection{Vertical iterative procedure}
\label{subsec:iterative procedure}

Iterative procedure is called for each band $\band{i}$ found in the preparatory phase within each band until constructed k-terms within each band will not satisfy prescribed precision. The procedure is applied to one species at a time. Since the species within each band can be ranked by their absorption strength, the iteration starts with the most optically dominant gas.
\begin{enumerate}[label=\roman*.]

    \item For the given band $\band{i}$ the reference calculations must be performed. We have then upward and downward radiative fluxes and cooling rates as functions of height: $F_{\downarrow}^{\text{LBL}}(z)$, $F_{\uparrow}^{\text{LBL}}(z)$ and $Q^{\text{LBL}}(z)$.

    \item For each atmospheric layer $l$, starting from the top, layer optical depth $\tau_{l}$ is retrieved so that the one-stream calculation of the layer-base downward flux matches the reference computed value for the same spectral subset. At each vertical step, \textit{effective cross-section} is obtained that mimics the absorption behavior in downward direction in a whole spectral interval $\band{i}$: 
    \begin{equation}
        \sigma_{\text{eff}, l}=\frac{1}{n\Delta z}\ln\frac{F_{\downarrow,l}}{F_{\downarrow,l-1}}, \; l=1,\ldots,H,
    \end{equation}
     where $H$ — number of atmospheric levels and $\Delta z$ is a thickness of one atmospheric layer, $n$ — species number density in proper units. By a simple interpolation, from the set of values $\sigma_{\text{eff}, l}$, a vertically-resolved effective cross-section $\sigma_{\text{eff}}(z)$ is constructed and utilized as an input for reference calculations in the next step. Note, that these effective cross-sections are internal auxiliary quantities used only during the construction of the k-distribution terms.

    \item Radiative transfer calculation is performed with effective absorption cross-section $\sigma_{\text{eff}}(z)$ in attempt to echo the reference line-by-line calculation for the whole band. As a result of that calculation, ``effective'' fluxes and cooling rates are computed: $F_{\downarrow, \text{eff}}(z)$, $F_{\uparrow, \text{eff}}(z)$ and $Q_{\text{eff}}(z)$.

    \item The resulting effective radiative fields are then compared with their line-by-line counterparts. By definition of $\sigma_{\text{eff}}(z)$, $F_{\downarrow, \text{eff}}(z)$ will be exactly equal to $F_{\downarrow}^{\text{LBL}}(z)$, while a difference might appear between $F_{\uparrow, \text{eff}}(z)$ vs $F_{\uparrow}^{\text{LBL}}(z)$ and $Q_{\text{eff}}(z)$ vs $Q^{\text{LBL}}(z)$. If the difference between these values remains within the prescribed precision, the iterative procedure is considered converged and can be terminated, proceeding to the next spectral interval $\band{i+1}$. If not — multiple k-terms must be introduced within the band, reflecting the division between weak and strong absorption (step 5).

    \item If the prescribed accuracy is not achieved, the spectral band is subdivided into subsets based on absorption strength. At a reference level $z^\ast$ we select a provisional subset $U_1$ of weakly absorbing wave numbers by imposing a threshold $S_1$ on the line-by-line absorption coefficients $\alpha_\nu(z^\ast)$,
    \[
        \nu \in U_1 \;\Longleftrightarrow\; \alpha_\nu(z^\ast) < S_1.
    \]
    Line-by-line reference fields $F_{\downarrow}^{\text{LBL}}(z)$, $F_{\uparrow}^{\text{LBL}}(z)$ and $Q^{\text{LBL}}(z)$ are then recomputed using only points in $U_1$, and the vertical inversion (steps~2–4) is applied to this subset to obtain an effective cross-section profile $\sigma_{\text{eff}}^{(1)}(z)$ and its associated effective fields $F_{\downarrow,\text{eff}}^{(1)}(z)$, $F_{\uparrow,\text{eff}}^{(1)}(z)$ and $Q_{\text{eff}}^{(1)}(z)$. If the errors in $F_{\uparrow}$ and $Q$ with respect to their line-by-line counterparts exceed the prescribed precision, $S_1$ is adjusted and the procedure is repeated until the widest acceptable subset $U_1$ is found. This subset defines the first $k$-term in the band; its wave numbers are removed, and the same procedure is iterated on the remaining points (with possibly a different $z^\ast$) to construct subsequent subsets $U_2, U_3, \ldots$ and their effective cross-sections $\sigma_{\text{eff}}^{(j)}(z)$.
    
\end{enumerate}

\subsection{Resulting k-table}

The resulting set of $k$-distribution terms is summarized in Table~\ref{tab:kterms}. Over the longwave interval 10–6000~cm$^{-1}$, we obtain $N = 16$ bands and only 32 k-terms. In terms of the number of radiative transfer solves, this parameterization is more compact than existing correlated-$k$ implementations for Venus. The scheme introduced by \textcite{takahashi2023development}, which adopts a very similar band layout (17 bands over 10–6425~cm$^{-1}$), requires about 3.5 times more solutions of the transport equations than our scheme. Other longwave parameterizations developed for Venus GCMs, such as the Net Exchange Rate–based scheme of \textcite{eymet2009net} later implemented by \textcite{lebonnois2015analysis}, employ considerably finer spectral and $k$-distribution discretizations and therefore involve a larger number of radiative transfer evaluations per call than the FKDM approach presented here.

For each band, we provide — in addition to the set of $M_i$ k-terms — a single representative spectral point $\tilde\nu_i^{\text{cld}}$ (Table~\ref{tab:kterms}) to be used for the computation of cloud optical properties. In the longwave spectral range considered here, cloud extinction and single-scattering properties vary only weakly across an individual band, so we assume them to be constant within the band and evaluate them at the prescribed $\tilde\nu_i^{\text{cld}}$. Additionally, as mentioned in section \ref{sec:code}, a band-averaged Planckian profile is introduced. Thus, each band $i$ is fully characterized by its number of k-terms $M_i$ and by its representative spectral point $\tilde\nu_i^{\text{cld}}$ and Planck function values, which together define the inputs required by the radiative transfer solver.

\begin{figure}[t]
  \centering
  \includegraphics[width=0.85\columnwidth]{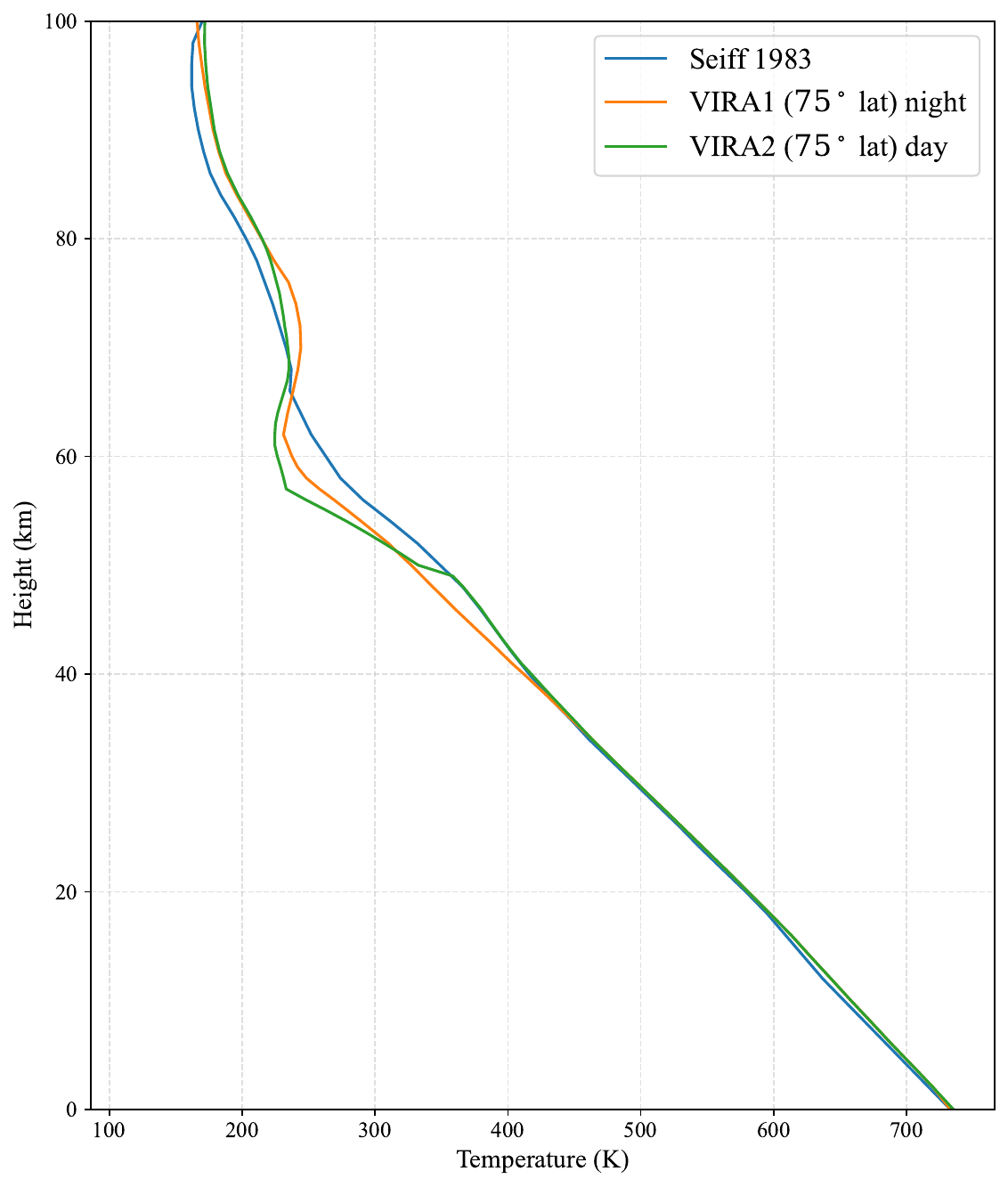}
  \caption{Representative Venus atmospheric temperature profiles (Seiff, 1983; VIRA-1 day- and night-side at $75^\circ$ latitude) used for the construction and sensitivity analysis of the FKDM $k$-distribution.}
  \label{fig:Tprofiles}
\end{figure}

The resulting $k$-distributions inherit the thermodynamic dependence of the underlying
line-by-line absorption cross-sections and therefore depend on temperature along the prescribed
pressure–temperature profiles. To quantify the sensitivity of the parameterization to
temperature, we repeated the construction of k-terms for two reference Venus profiles
that reflect typical hemispheric conditions below 100 km (see Fig.~\ref{fig:Tprofiles}).
Below about 100 km ($p \gtrsim 10^{-5}$ atm), the profiles remain within roughly 20 K of each other.

\begin{table}[ht]
\centering
\caption{$k$-distribution table}
\label{tab:kterms}
\small
\setlength{\tabcolsep}{3pt}
\begin{tabular}{c c c c c}
\hline
Band & Spectral region (cm$^{-1}$) & Active species & $\tilde\nu_i^{\text{cld}}$ (cm$^{-1}$) & $M_i$ \\
\hline
1  & 10--200    & H$_2$O         & 100  & 2  \\
2  & 200--400   & H$_2$O         & 300  & 2  \\
3  & 400--500   & CO$_2$         & 450  & 1  \\
4  & 500--600   & CO$_2$         & 550  & 1  \\
5  & 600--760   & CO$_2$         & 680  & 10 \\
6  & 760--900   & CO$_2$         & 830  & 2  \\
7  & 900--1100  & CO$_2$         & 1000 & 5  \\
8  & 1100--1210 & CO$_2$, SO$_2$ & 1150 & 1  \\
9  & 1210--1300 & CO$_2$, SO$_2$ & 1250 & 1  \\
10 & 1300--1400 & CO$_2$, SO$_2$ & 1350 & 1  \\
11 & 1400--1800 & H$_2$O, CO$_2$, SO$_2$ & 1600 & 1 \\
12 & 1800--2650 & CO$_2$         & 2000 & 1  \\
13 & 2650--3000 & CO$_2$         & 2800 & 1  \\
14 & 3000--4100 & CO$_2$         & 3500 & 1  \\
15 & 4100--4400 & H$_2$O, CO$_2$ & 4300 & 1  \\
16 & 4400--6000 & CO$_2$         & 5000 & 1  \\
\hline
\multicolumn{4}{l}{Total $N=16$} & 32 \\
\hline
\end{tabular}
\end{table}

For the second $k$-term in the 600--760~cm$^{-1}$ band, we compared the corresponding logarithms of effective cross sections, $S=\ln\sigma_{\mathrm{eff}}$, obtained for different representative Venus temperature profiles as functions of $w=\ln p$. In the interval $w\in(-4,7)$, corresponding to altitudes $z\in(100,50)$~km, the resulting $S(w)$ profiles remain close to each other, reflecting the modest temperature differences within this altitude range. A similar behavior was found for all other bands, indicating weak sensitivity of the effective cross sections to realistic variations in the Venus thermal structure. Based on that, a single ``baseline'' effective cross-section profile $S^{\ast}(w)$ can be adopted for all considered temperature profiles, except in the 600--760 and 760--900~cm$^{-1}$ intervals.
For these two bands we retain an explicit temperature dependence by applying, at each
level $w$, an empirical correction
\begin{equation}
S\!\left(T(w)\right)
=
S^{\ast}\!\left(T^{\ast}(w)\right)
+
\alpha(w)\,
\frac{T(w)-T^{\ast}(w)}{T^{\ast}(w)},
\label{eq:temp_correction}
\end{equation}
where $T^{\ast}(w)$ is the baseline temperature profile and $\alpha(w)$ is determined from
numerical experiments. In the final parameterization, the functions $S^{\ast}(w)$ and
$\alpha(w)$ are provided in tabular form within the Fortran code and are accessed internally by the FKDM module described in section \ref{sec:code}.

\begin{figure}
  \centering
    \includegraphics[width=0.85\columnwidth]{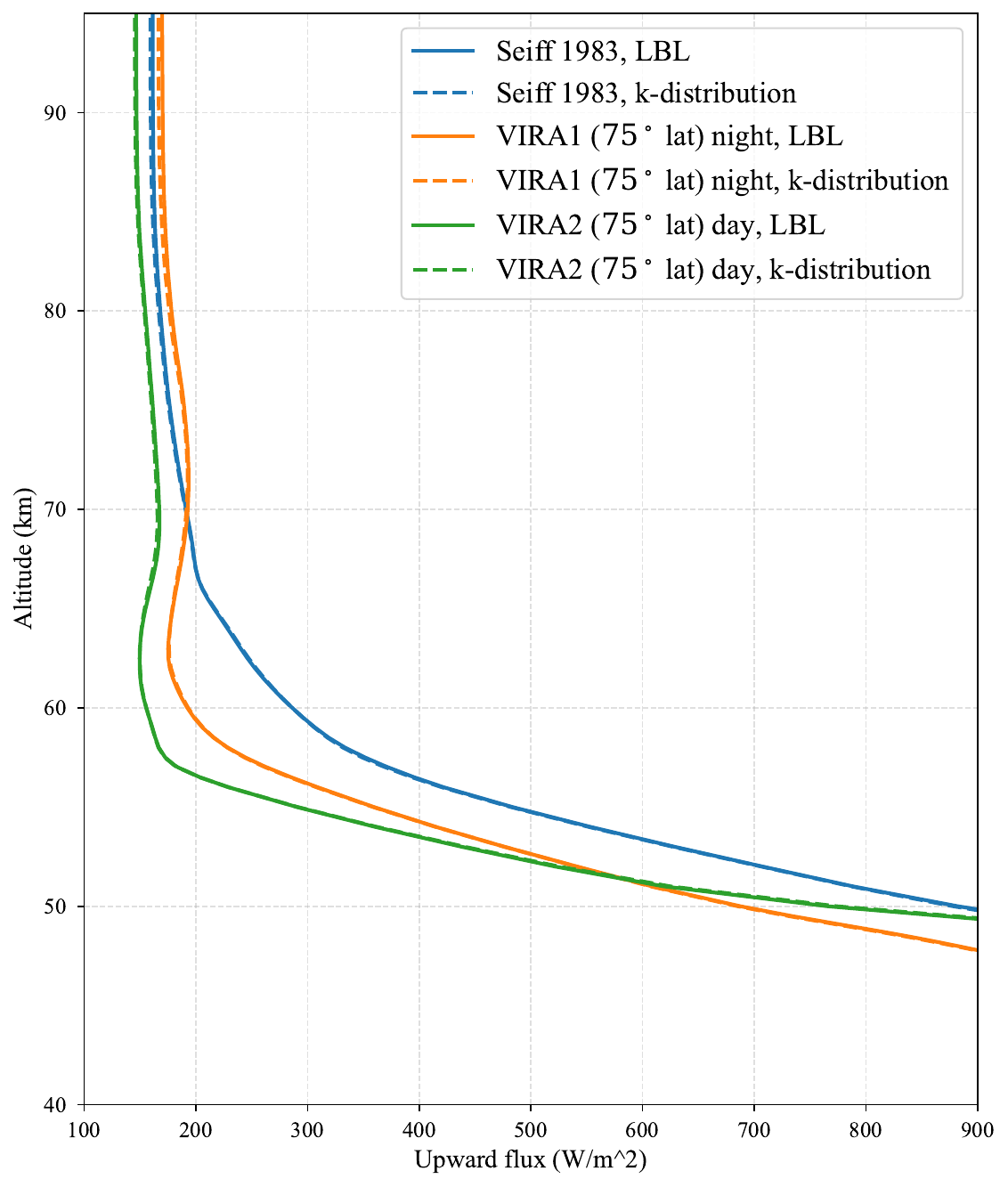}
    \caption{Vertical profiles of upward longwave radiative fluxes computed with the FKDM $k$-distribution scheme (dashed lines) and the line-by-line Monte-Carlo reference model (solid lines) for several representative Venus atmospheric profiles. 
}
  \label{fig:Fup}
\end{figure}

\begin{figure}
  \centering
    \includegraphics[width=0.85\columnwidth]{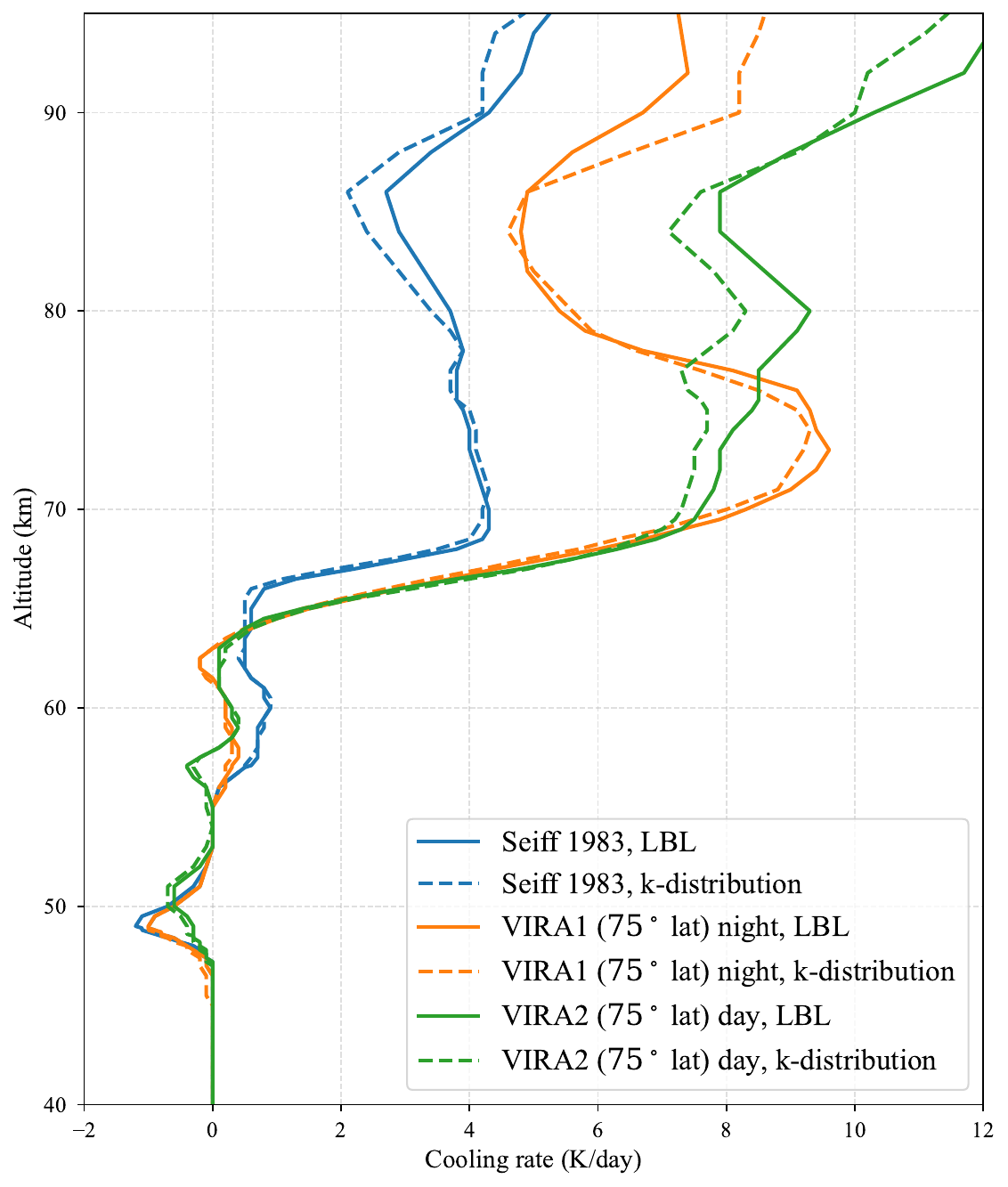}
    \caption{Comparison of radiative cooling rate profiles obtained with the FKDM $k$-distribution scheme (dashed lines) and the line-by-line Monte-Carlo reference calculations (solid lines) for the same set of Venus atmospheric profiles as in Fig.~\ref{fig:Fup}.
}
  \label{fig:cooling rates}
\end{figure}

Figures~\ref{fig:Fup} and~\ref{fig:cooling rates} compare upward radiative fluxes and cooling rates computed with the FKDM parameterization against the corresponding line-by-line reference calculations for representative Venus atmospheric profiles shown in Fig.~\ref{fig:Tprofiles}. Close agreement between the two solutions is evident for the upward fluxes and is of <2\% below 95 km. As a basic consistency check, the globally averaged outgoing thermal flux at TOA is expected to be $157 \pm 6~\mathrm{W\,m^{-2}}$ \parencite{titov2007radiation, eymet2009net}, corresponding to about 3.8~\% of plausible deviation. The sensitivity study by \textcite{Haus2015} reports that uncertainties in cooling rates below 95 km are about 0.25 $\mathrm{K\,day^{-1}}$. In our implementation, discrepancies do not exceed approximately 1.5 $\mathrm{K\,day^{-1}}$ below 95 km, decreasing to below 1.2 $\mathrm{K\,day^{-1}}$ below 90 km and to 0.3 $\mathrm{K\,day^{-1}}$ below 75 km approaching the level of uncertainties inherent to reference calculations.

\section{Radiative transfer model}
\label{sec:modeling inputs}

As mentioned in the previous section, reference radiative transfer modeling is an inherent part of the adopted $k$-distribution technique. Line-by-line reference modeling of the Venus atmosphere—performed in practice as high-resolution quasi-monochromatic calculations with a precise solver over a broad spectral range—is itself a non-trivial problem, and only a few such models have been documented. One prominent example is the research conducted by \textcite{Haus2015,Haus2016}, who couple a discrete-ordinate solver with absorption cross-sections precomputed by a line-by-line procedure, to compute fluxes and heating/cooling rates over $0.125$–$1000~\mu\mathrm{m}$ using an optimized spectral grid. \textcite{lee2016sensitivity} use a fast line-by-line SHDOM-based model over \(50\)–\(8300~\mathrm{cm}^{-1}\) (\(1.2\)–\(200~\mu\mathrm{m}\)) at \(0.1~\mathrm{cm}^{-1}\) resolution to quantify the influence of CO$_2$ collision-induced absorption and trace gases (SO$_2$, H$_2$O, OCS) on the Venus net thermal flux and to match Pioneer Venus probe observations.
 Other broadband multiple-scattering models for Venus include the discrete-ordinate RTM of \textcite{lee2011discrete}, spanning $0.1$–$260~\mu\mathrm{m}$ and used both for forward simulations and radiative–convective equilibrium studies, as well as parameterized infrared schemes based on net-exchange-rate or correlated-$k$ formalisms that are currently embedded in Venus GCMs \parencite{eymet2009net, lebonnois2015analysis, mendonca2015new}. Recently, \textcite{takahashi2023development} have developed a unified line-by-line and correlated-$k$ radiation framework for CO$_2$-dominated atmospheres (Venus and Mars), in which a high-resolution line-by-line model is used to generate and validate $k$-tables for a generalized two-stream solver and associated radiative–convective models. In the 1–5~$\mu\mathrm{m}$ spectral interval, \textcite{ojha2025near} employed a LIDORT-based multiple-scattering line-by-line model, aimed at supporting surface emissivity retrievals for forthcoming Venus orbiter missions.

We employed a reference line-by-line radiative transfer model based on the Monte-Carlo method \parencite{liou2002introduction} and its algorithmic implementation developed in \textcite{fomin1998model,fomin2006monte}. The same code family has been used previously in our FKDM $k$-distribution study for the UV region \parencite{Fomin2022}, and the Earth version has participated in radiation-code intercomparisons for climate applications \parencite{Oreopoulos2012continual}. The model provides an accurate treatment of gaseous absorption and particulate multiple scattering for longwave radiation in plane-parallel slab atmospheres with one or several cloud and aerosol layers. Multiple scattering is simulated explicitly, and the residual error in fluxes and cooling rates decreases as $K^{-1/2}$ with the total number of photon histories $K$, independently of the number of spectral points \parencite{fomin2006monte}. This statistical convergence, together with the fact that the error is governed by photon statistics rather than by the spectral grid, makes the approach particularly suitable for broadband longwave calculations in highly scattering, optically thick cloud layers \parencite{fomin2006monte}. Consequently, in the present work we simply adopt the native high-resolution grid at which monochromatic absorption coefficients are precomputed, with a spacing of $1/2048~\mathrm{cm}^{-1}$, letting the Monte-Carlo statistics control the overall accuracy of the reference solution. 

We present the resulting $k$-distribution set for the altitude range of 0-100 km. Choosing this altitude range largely depends upon the altitude limit for reliable reference modeling with existing thermal profiles. Temperature structure in 100-150 km region remains observationally underconstrained, while obtained temperature change rates can have substantial uncertainties even in 90-100 km range as mentioned by \textcite{Haus2015,Haus2016}. Only three gaseous species are treated as main absorbers in the modeling atmosphere: CO$_2$, H$_2$O, and SO$_2$. Trace gases (e.g. CO, OCS, HCl, HF) are omitted from the generation of the $k$-distribution, as their primary importance lies in spectroscopic details in narrow windows rather than in energetic studies \parencite{Haus2015,Haus2016} or in applications for GCMs \parencite{mendonca2015new}.

The FKDM method assumes that k-terms are generated for a clear-sky, purely absorbing atmosphere. Validation is then performed in a full atmosphere that includes clouds and scattering. Because the reference model is used solely for validation, only certain inputs must be specified with high accuracy. The most critical are the monochromatic absorption cross-sections (or absorption coefficients), since these directly determine the true absorption behavior against which the k-terms are tested. The temperature profile must also be chosen carefully, both to represent the Venus P–T environment and to capture the correct temperature dependence of the molecular absorption. Other inputs—such as the cloud particle size distributions—need only be realistic rather than exact, providing conditions representative of the Venus atmosphere for the purposes of validation.

\subsection {Atmospheric profiles}
We utilized a set of three representative and realistically contrasting atmospheric profiles: the low-latitude midnight profile introduced by \textcite{Seiff1983}, VIRA-1 midnight 75$^\circ$ latitude profile from \textcite{Seiff1985} and a combined profile with VIRA-2 75$^\circ$ latitude day-side profile introduced by \textcite{Zasova2006} for solar longitude sector $L_s=20^\circ-90^\circ$ in the middle atmosphere 50-100 km and relevant VIRA-1 profile below 50 km. Temperature profiles are depicted on Fig. \ref{fig:Tprofiles}.

Initial VIRA-like data provide only total column abundance, so to decouple individual densities, volume mixing rations of each species must be introduced. The volume mixing ratio of CO$_2$ is assumed to be constant with altitude and is fixed at 0.965. H$_2$O abundance is set to 32.5 ppmv below 50 km, consistent with near-IR determinations of $\sim$30–35 ppmv in the lower atmosphere \parencite{Arney2014,Fedorova2016}, and to 3 ppmv near and above the cloud tops ($\sim$70 km), consistent with nadir and occultation measurements of $\sim$1–3 ppmv \parencite{Cottini2012,Fedorova2016}. SO$_2$ abundances are taken from Fig.5 of model input by \textcite{Haus2015}, that describe a vertically uniform lower-atmosphere mixture below 50 km, with volume mixing ratios of 32.5 and 150 ppmv, respectively. 

\subsection {Absorption by gaseous components}
\label{sec:absorption by gaseous}

Our recently introduced MARFA code \parencite{Razumovskiy2025MARFA} is employed for calculation of absorption coefficients based on line parameters. MARFA computes absorption coefficients at a high spectral resolution of \(1/2048~\mathrm{cm^{-1}}\), which is maintained into the far wings of spectral lines through the use of an efficient multi-grid interpolation algorithm~\parencite{fomin1995effective}. Particular emphasis is placed on evaluating line shapes with large cut-offs, an essential feature in presence of uncertainties in spectroscopic parameters and continuum absorption in Venus modeling. In the absence of a dedicated, fully consistent Venus spectroscopic database, line parameters must be assembled from several sources. For spectroscopic input, we utilized the HITEMP database \parencite{rothman2010HITEMP} for CO$_2$ and H$_2$O, which provides comprehensive high-temperature line lists essential for Venus' broad thermal range. For SO$_2$, we relied on HITRAN2020 database \parencite{Gordon2022}.

The choice of continuum absorption parameterization in Venus radiative transfer modeling depends on the intended application. Pressure-induced continuum has a non-negligible influence in retrieval studies, for example on cloud opacity and related cloud parameters \parencite{Haus2013selfconsistent, evdokimova2025cloud}, or in simulations of deep-atmosphere net fluxes for comparison with entry probe measurements \parencite{lee2016sensitivity}. In such studies—typically restricted to relatively narrow spectral intervals—continuum absorption is often described either by physically based laboratory determinations  \parencite{moskalenko1979pressure, gruszka1997roto, richard2012new, stefani2013experimental, tran2024collision} or by introducing additional empirical opacity to match observations in near-infrared windows \parencite{bezard1990chi-factor, de1995water, marcq2006remote, Bezard2011}. A sensitivity analysis by \textcite{Haus2015} demonstrated that, for the purpose of computing broadband net fluxes and heating rates, continuum absorption can be safely neglected; this conclusion was further supported by their energy-balance study \parencite{Haus2016}. In those studies, the relevant collisional line-mixing and pressure-induced effects were incorporated implicitly through sub-Lorentzian $\chi$-factors and far-wing truncation. Following the recommendations of \textcite{Haus2015,Haus2016}, we similarly applied the sub-Lorentzian correction of \textcite{Tonkov1996} to the CO$_2$ far wings and truncated lines at 250 cm$^{-1}$. No intensity cut-off was imposed, as MARFA efficiently treats the full line list. Because the primary purpose of our $k$-distribution tables is to enhance the computational efficiency of radiative transfer modules in Venus climate modeling (where emphasis is made on broad spectral ranges), adopting a Haus-type treatment of continuum absorption \parencite{Haus2015,Haus2016} is well justified. In future, our line-by-line Monte-Carlo model will be released as a standalone tool suitable for a broader class of radiative transfer problems and we plan to incorporate a physically based continuum absorption model.

In a recent study, \textcite{chaverot2025spect} introduced updated $\chi$-factor parameterizations for several planetary environments, including Venus, based on a comprehensive compilation and re-analysis of laboratory data. In their framework, the definition of continuum absorption is made more rigorous by adopting a fixed line-wing cut-off at 25~cm$^{-1}$ from the line center. The same $\chi$-factor formalism is then used to construct separate CO$_2$ line-wing continua over 25–1500~cm$^{-1}$. Additionally, dedicated continuum files, accounting for collision-induced and dimer absorption, are provided. In future versions of our $k$-distribution scheme (e.g. for short-wave radiation) we plan to adopt these new $\chi$-factors and continuum absorption parameterizations.

\subsection {Cloud aerosol properties}
Cloud particle concentrations are adopted from the initial cloud model introduced by \textcite{Haus2015}, which is based on the VIRTIS-derived self-consistent retrievals from \textcite{Haus2013selfconsistent}. Per-mode concentrations follow the analytical expressions provided in Table 1 of their study \parencite{Haus2015}. Log-normal size distribution parameters are likewise drawn from \textcite{Pollack1993}. The model incorporates four modes of spherical liquid droplets, each composed of a sulfuric acid-water mixture (H$_2$SO$_4$-H$_2$O) with a uniform concentration of 75 wt\% H$_2$SO$_4$ (by weight) across all modes and altitudes. Refractive index data are sourced from the study by \textcite{Palmer1975}. Based on Mie scattering theory \parencite{wiscombe1980improved}, we employed the same computational code as in our prior work \parencite{Fomin2022} to derive the cloud microphysical optical properties. These parameters serve as inputs for our radiative transfer model. Over the 10–6000~cm$^{-1}$ spectral interval considered in the present study, cloud droplets behave as strong absorbers throughout most of the thermal infrared ($\lesssim 3500$~cm$^{-1}$), approaching nearly conservative scattering only in the short-wave near-IR tail ($\gtrsim 4000$~cm$^{-1}$) \parencite{Haus2015}.

\section{Code and data}
\label{sec:code}

To support the use of the longwave FKDM parameterization in radiative transfer models and general circulation models, we provide a compact Fortran implementation in a public repository\footnote{\url{https://github.com/Razumovskyy/Venus-longwave-fkdm}}. The repository contains a Fortran driver which computes vertical profiles of $k$-distribution terms and band-integrated Planck functions for arbitrary atmospheric columns. The code is organized as a Fortran Package Manager (fpm) project, such that compilation and execution require only \texttt{fpm run}. By default, two diagnostic outputs are produced: (i), containing the volume absorption coefficients $k_{i,j}(z)$, and (ii), containing the corresponding band-integrated Planck functions. These reproduce the quantities a host GCM or radiative transfer solver would typically obtain through direct calls to the FKDM routines. The implementation is modular and designed to operate either as a stand-alone tool for one-dimensional radiative transfer tests or as transferable module component for any Venus GCM. The repository is equipped with a \texttt{Readme} file to facilitate onboarding.

In operational mode, a host model supplies vertical profiles of pressure, temperature, and mixing ratios of CO$_2$, H$_2$O, and SO$_2$, which are passed to the FKDM module. For each longwave band and $k$-term (16 bands, 32 channels), the module returns the volume absorption coefficients $k_{i,j}(z)$ and the associated band-integrated Planck function values $B_i(z)$ ($i$ – index counting band, $j$ — index counting channel within the band). The radiative transfer solver then evaluates the fluxes and cooling rates using its preferred method (e.g., two-stream, discrete ordinate, etc.) and aggregates them over all bands and $k$-terms. Cloud optical properties are incorporated in the usual way by evaluating extinction and single-scattering parameters at the representative cloud wavenumbers $\tilde{\nu}_i^{\mathrm{cld}}$ for each band listed in Table \ref{tab:kterms}.

From the user's perspective the FKDM module essentially acts as a black box and provides a lightweight mapping:  
\[
\begin{aligned}
\bigl(P(z),\, T(z),\, q_{\mathrm{CO_2}}(z),\, q_{\mathrm{H_2O}}(z),\, q_{\mathrm{SO_2}}(z)\bigr)
\\[0.3em]
\longrightarrow\quad
\{k_{i,j}(z),\, B_i(z)\},
\end{aligned}
\]

where $q$ is mixing ratio. It can be called directly from the host model’s radiation scheme with minimal changes to the existing infrastructure.

\section{Conclusion}
\label{sec:conclusion}
We have developed a minimal-set $k$-distribution parameterization for longwave gaseous absorption in the Venus atmosphere and validated it against a high-resolution line-by-line Monte-Carlo radiative transfer model. The FKDM approach produces only 32 spectral channels across 16 bands for CO$_2$, H$_2$O and SO$_2$, while maintaining accuracy comparable to standard correlated-$k$ implementations and potentially beyond that. The method natively accounts for vertical inhomogeneity, avoids assumptions about inter-level correlation of monochromatic absorption, and allows direct control over the accuracy–efficiency balance during construction of the k-terms. Although developed and validated here for Venus case, the method is general and can be adapted to other planetary regimes.

Validation across a set of representative Venus temperature profiles demonstrates that upward radiative fluxes computed with the FKDM scheme closely reproduce the corresponding line-by-line results up to 95 km, reflecting the flux-matching strategy employed during the construction of the $k$-distribution terms. Cooling rates are reproduced with acceptable accuracy throughout the lower and middle atmosphere, remaining consistent with line-by-line calculations up to 90 km. The compactness of the resulting parameterization yields a substantial reduction in radiative transfer evaluations, making the FKDM approach well suited for Venus GCMs and radiative–convective models.

A FKDM Fortran module is made publicly available to ensure reproducibility and to facilitate integration into existing radiative transfer schemes. Future work will extend the methodology to the shortwave spectral range and incorporate updated spectroscopic treatments, including up-to-date $\chi$-factors and continuum parameterizations, as they become available.
\newpage

\bibliographystyle{elsarticle-harv}
\bibliography{references}

\end{document}